%% file: main.tex
\RequirePackage{lineno}
\documentclass[prd,twocolumn,showpacs,amsmath,amssymb,nofootinbib]{revtex4-2}
\usepackage{overpic}
\usepackage{graphicx}
\usepackage{epsfig}
\usepackage{dcolumn}
\usepackage{bm}
\usepackage{rotating}
\usepackage{multirow}
\usepackage{booktabs}
\usepackage{appendix}
\usepackage{subfigure}
\usepackage{hhline}
\usepackage{amssymb}
\usepackage{lineno}
\usepackage{hyperref}
\usepackage{mathrsfs}
\usepackage{verbatim}
\usepackage{amsmath}
\usepackage{amssymb}
\usepackage{amsfonts}
\usepackage{amsbsy}
\usepackage{float}

\hypersetup{colorlinks,linkcolor=blue,anchorcolor=blue,citecolor=blue}
\include{def-com}

\let\oldequation\equation
\let\oldendequation\endequation

\renewenvironment{equation}
  {\linenomathNonumbers\oldequation}
  {\oldendequation\endlinenomath}

\begin{document}

\title{\boldmath Search for $\eta_c(2S)\to p\bar{p}$ and branching fraction measurements of $\chi_{cJ} \to p\bar{p}$ via $\psi(2S)$ radiative decays}
\input{authorlist_2024-08-05}


\begin{abstract}

Using $(27.12\pm0.14) \times 10^{8}$ $\psi(2S)$ events collected by
the BESIII detector operating at BEPCII, we search for the decay
$\eta_c(2S)\to p\bar{p}$ via the process $\psi(2S)\to
\gamma\eta_c(2S)$, and only find a signal with a significance of
$1.7\,\sigma$. The upper limit of the product branching fraction at
the 90\% confidence level is determined to be $\mathcal{B}(\psi(2S)\to
\gamma\eta_c(2S))\times \mathcal{B}(\eta_c(2S)\to p\bar{p})<2.4\times
10^{-7}$. The branching fractions of $\chi_{cJ}\to p\bar{p}~(J=0,1,2)$
are also measured to be $\mathcal{B}(\chi_{c0}\to
p\bar{p})=(2.51\pm0.02\pm0.08)\times 10^{-4}$,
$\mathcal{B}(\chi_{c1}\to p\bar{p})=(8.16\pm0.09\pm0.25)\times
10^{-4}$, and $\mathcal{B}(\chi_{c2}\to
p\bar{p})=(8.33\pm0.09\pm0.22)\times 10^{-4}$, where the first
uncertainty is statistical and the second systematic.

\end{abstract}

\maketitle
\oddsidemargin -0.2cm
\evensidemargin -0.2cm

\section{\boldmath Introduction}

Experimental and theoretical studies of charmonium states play an
important role in understanding Quantum Chromodynamics (QCD). Since
the first member of this family, the $J/\psi$, was observed in
experiment~\cite{jpsi_1,jpsi_2,jpsi_3}, other charmonium states below
the open-charm threshold have been discovered. Among these states, the
$\eta_{c}(2S)$ and $h_{c}(1P)$ are less well understood. The spin-singlet 
state $\eta_{c}(2S)$ was first observed by the Belle
experiment in $B$ meson decay $B \to K\eta_{c}(2S)$, via $\eta_{c}(2S)
\to K_{S}^{0}K^{\mp}\pi^{\pm}$~\cite{etacp_Belle}. This state was
subsequently confirmed in several
experiments~\cite{etacp_2,etacp_3,etacp_4,etacp_5}. In 2012, BESIII
observed $\eta_{c}(2S)$ in the radiative transition $\psi(2S) \to
\gamma\eta_{c}(2S)$, where $\eta_{c}(2S)$ is reconstructed by
$K_{S}^{0}K^{\pm}\pi^{\mp}$ and $K^+K^-\pi^0$ final
states~\cite{etacp_BESIII}. Our understanding of $\eta_{c}(2S)$ decay
modes is still limited. To date, only seven decay modes of
$\eta_{c}(2S)$ have been observed experimentally, with the largest
branching fraction of $(1.9\pm1.2)\%$ for the $K\bar{K}\pi$
mode~\cite{PDG}.

Among the various decay channels, the decay of $\eta_c(2S)$ into a
proton-antiproton ($p\bar{p}$) pair has attracted particular
interest. In 2013, the decay $\eta_c(2S) \to p\bar{p}$ was searched
for using $106\times10^6$ $\psi(2S)$ events collected by the BESIII
detector~\cite{besiii2013}. The statistical significance of the
$\eta_{c}(2S)$ signal was found to be $1.7\,\sigma$, and the upper
limit of the product branching fraction $\mathcal{B}(\psi(2S)\to
\gamma\eta_c(2S)) \times \mathcal{B}(\eta_c(2S)\to p\bar{p})$ at the 90\%
confidence level~(C.~L.) determined to be $1.4\times10^{-6}$. The
first observation of $\eta_c(2S) \to p\bar{p}$ was reported by the
LHCb experiment with a statistical significance of $6.4\,\sigma$,
where the $\eta_c(2S)$ resonance is produced in the decay $B^+
\to[c\bar c]K^+$. The product branching fraction normalized to the
$J/\psi$ intermediate state is given as $\frac{\mathcal{B}(\psi(2S)\to
\eta_c(2S) K^+ ) \times \mathcal{B}(\eta_c(2S) \to
p\bar{p})}{\mathcal{B}(\psi(2S)\to J/\psi K^+) \times
\mathcal{B}(J/\psi \to
p\bar{p})}=(1.58\pm0.33\pm0.09)\times10^{-2}$~\cite{LHCb}.

 Theoretically, S.~J.~Brodsky and G.~P.~Lepage~\cite{helicity_1981}
 predict that total hadron helicity is conserved in large
 momentum transfer processes, implying the decays of
 $\eta_c(1S)/\chi_{c0}/h_c/\eta_c(2S)$ to $p\bar{p}$ are forbidden by
 the helicity selection rule in massless QCD models. Another topic of
 interest in charmonium decays is the branching fraction ratio. As
 spin-singlet partners of $\psi(2S)$ and $J/\psi$, the $\eta_{c}(2S)$ and
 $\eta_{c}(1S)$, can decay into light hadrons similarly. Anselmino
 {\it et al.}~\cite{Ansel} assumed for all unforbidden hadronic
 channels that $\frac{\mathcal{B}(\eta_c(2S) \to
 h)}{\mathcal{B}(\eta_c(1S) \to h)} \approx \frac{\mathcal{B}(\psi(2S)
 \to h)}{\mathcal{B}(J/\psi \to h)} \approx 0.128$. However,
 K.~T.~Chao {\it et al.}~\cite{Chao} argue that this ratio should be
 $\frac{\mathcal{B}(\eta_c(2S) \to h)}{\mathcal{B}(\eta_c(1S) \to h)}
 \approx 1$, or $1/2$ if there is a mixture with a glueball. Using known
 branching fractions of $\eta_c(2S)$ and $\eta_c(1S)$, a global fit is
 performed and experimental results are found to be significantly different
 from the above theoretical predictions~\cite{Wanghongpeng}.

In this paper, using $(27.12\pm0.14)\times10^8$ $\psi(2S)$ events
collected by the BESIII detector in 2009, 2012, and
2021~\cite{number}, the decay $\eta_c(2S) \to p\bar{p}$ is searched for
through the radiative transition $\psi(2S)\to \gamma \eta_c(2S)$. However,
no significant signal is observed.  With the same analysis strategy,
the branching fractions of $\chi_{cJ}\to p\bar{p}~(J=0,1,2)$ are
determined with improved precision.

\section{\boldmath BESIII Detector and Monte Carlo Simulation}

The BESIII detector~\cite{bes3:detector_a} records symmetric $e^+e^-$
collisions provided by the BEPCII storage ring~\cite{bes3:detector_b}
in the center-of-mass energy ranging from 1.85 to 4.95~GeV, with a
peak luminosity of $1.1\times10^{33}$~cm$^{-2}$s$^{-1}$ achieved at
$\sqrt{s}=3.773$ GeV.  BESIII has collected large data samples in this
energy region~\cite{bes3:detector_c}. The cylindrical core of the
BESIII detector covers 93\% of the full solid angle and consists of a
helium-based multilayer drift chamber~(MDC), a time-of-flight
system~(TOF), and a CsI(Tl) electromagnetic calorimeter~(EMC), which
are all enclosed in a superconducting solenoidal magnet providing a
1.0~T magnetic field. The solenoid is supported by an octagonal
flux-return yoke with resistive plate counter muon identification
modules interleaved with steel. The charged-particle momentum
resolution at $1~{\rm GeV}/c$ is $0.5\%$, and the resolution of the
specific ionization energy (d$E$/d$x$) is $6\%$ for electrons from
Bhabha scattering. The EMC measures photon energies with a resolution
of $2.5\%$ ($5\%$) at $1$~GeV in the barrel (end-cap) region. The time
resolution in the TOF plastic scintillator barrel region is 68~ps,
while that in the end-cap region was 110~ps. The end-cap TOF system was
upgraded in 2015 using multi-gap resistive plate chamber technology,
providing a time resolution of 60~ps, which benefits 85\% of the data
used in this analysis~\cite{etof}.

Monte Carlo (MC) simulated samples produced with  {\sc
geant4}-based~\cite{bes3:geant4} software, which includes the
geometric description of the BESIII detector and the detector
response, are used to determine detection efficiencies and to estimate
backgrounds. The simulation models the beam energy spread and initial
state radiation (ISR) in the $e^+e^-$ annihilations with the generator
{\sc kkmc}~\cite{bes3:kkmc}. The inclusive MC sample includes the
production of the $\psi(2S)$ resonance, the ISR production of the
$J/\psi$, and the continuum processes incorporated in {\sc
kkmc}~\cite{bes3:kkmc}. Known decay modes are modeled with {\sc
evtgen}~\cite{bes3:evtgen1,bes3:evtgen2} using branching fractions
taken from the Particle Data Group (PDG)~\cite{PDG}. The remaining
unknown charmonium decays are modeled with {\sc
lundcharm}~\cite{bes3:lundcharm1,bes3:lundcharm2}.  Final state
radiation~(FSR) from charged final state particles is incorporated
using {\sc photos}~\cite{bes3:photos}. Exclusive MC
samples are generated to determine the detection efficiency and optimize
selection criteria. The process of $\psi(2S) \to \gamma
\chi_{cJ}/\eta_{c}(2S)$ is generated following the angular
distribution of ($1+\lambda\cos^2\theta_1$), where $\theta_1$ is the
polar angle of the radiative photon in the rest frame of $\psi(2S)$,
and the value of $\lambda$ is set to be 1 for $\eta_{c}(2S)$, and 1,
-1/3, 1/13 for $\chi_{cJ}~(J=0,1,2)$, respectively~\cite{p2gcj}. The
$\chi_{cJ} \to p\bar{p}$ decays are generated with
($1+\alpha \cos^2\theta_2$) distribution, where $\theta_2$ is the
polar angle of proton in the $\chi_{cJ}$ helicity frame, and $\alpha$
is determined from data. The $\eta_{c}(2S) \to p\bar{p}$ decay is
generated with a phase-space~(PHSP) model.

\section{\boldmath Event Selection}
\label{sec:selection}

The final state of interest contains two charged particles and one
neutral particle. Charged tracks detected in the MDC are required to
be within a polar angle ($\theta$) range of $|\cos\theta|<0.93$, where
$\theta$ is defined with respect to the $z$-axis, the symmetry axis of 
the MDC. The distance of the closest approach to the
interaction point (IP) must be less than 10~cm along the $z$-axis,
$|V_{z}|$, and less than 1~cm in the transverse plane, $|V_{xy}|$. Two
good charged tracks are required in the final state, and the total
charge must be equal to zero.

The particle identification (PID) for charged tracks combines
measurements of the d$E$/d$x$ in the MDC and the flight time in the
TOF to form likelihoods $\mathcal{L}(h)$ $(h=p, K, \pi)$ for each
hadron ($h$) hypothesis. A charged track is identified as a proton
when the proton hypothesis has the maximum likelihood value,
i.e. $\mathcal{L}(p)>\mathcal{L}(K)$ and
$\mathcal{L}(p)>\mathcal{L}(\pi)$. The two good charged tracks must be
identified as a proton and an anti-proton.

In the selection of good photon candidates, the deposited energy for a
cluster is required to be larger than 25~MeV in both the barrel ($ |\cos
\theta| < 0.80 $) and end-cap ($ 0.86 < |\cos \theta| < 0.92$)
regions.  To suppress electronic noise and unrelated showers, 
the difference between the EMC time and the event start time is
required to be within [0, 700]~ns. The opening angle between the
cluster and the closest good charged track is required to be larger
than $20^{\circ}$ for a proton and $30^{\circ}$ for an anti-proton.  The
number of good photon candidates is required to be greater than zero.

A vertex fit of the two charged tracks is performed to check if they
are consistent with coming from the IP. Next, a four-constraint (4C)
kinematic fit~\cite{4c} is performed with all the final state
particles, where the summed four-momentum of two charged tracks and a
neutral track is constrained to the initial four-momentum of
$\psi(2S)$ . For the events with more than one photon candidate, the
photon with the minimum $\chi^2_{4\rm C}$ value is selected. The $\chi_{4\rm C}^2$ is required
to be less than 60, which is optimized by maximizing the
figure-of-merit defined as $S/\sqrt{S+B}$, where $S$ and $B$ are the
expected yields of signal and background events in the $\eta_{c}(2S)$
signal region normalized to data. $S$ is estimated by
$N_{\psi(2S)}^{\rm tot} \times \mathcal{B}(\psi(2S) \to \gamma
\eta_{c}(2S)) \times \mathcal{B}(\eta_{c}(2S) \to p\bar{p}) \times
\epsilon^{\rm MC}$, where $N_{\psi(2S)}^{\rm tot}$ is the number
of $\psi(2S)$ events, $\mathcal{B}(\psi(2S) \to \gamma \eta_{c}(2S))$
is taken from PDG~\cite{PDG}, $\mathcal{B}(\eta_{c}(2S) \to p\bar{p})$
is set to the LHCb measurement ~\cite{LHCb}, and $\epsilon^{\rm
MC}$ is the detection efficiency. $B$ is the number of background
events estimated from the inclusive MC sample.

\section{\boldmath Background Analysis} 

The inclusive MC sample shows that the main backgrounds are
from $\psi(2S) \to \gamma p\bar{p}$, $p\bar{p}$, and $\pi^{0}
p\bar{p}$ processes. The other backgrounds account for only 1\% of all
the background events, which is thereby negligible, and there is no
peaking background. The non-resonant $\psi(2S) \to \gamma p\bar{p}$
background shares the same final state as the signal channel and
therefore cannot be suppressed. The other two backgrounds $\psi(2S)
\to p\bar{p}$ and $\psi(2S) \to \pi^{0} p\bar{p}$, which have one less
or one more photon, respectively, as well as the contribution from the 
continuum production will be discussed in detail below.


\subsection{\boldmath Background of $\psi(2S) \to p\bar{p}$ } 

\begin{figure}[htbp] \begin{center}
\begin{overpic}[width=0.48\textwidth,angle=0]{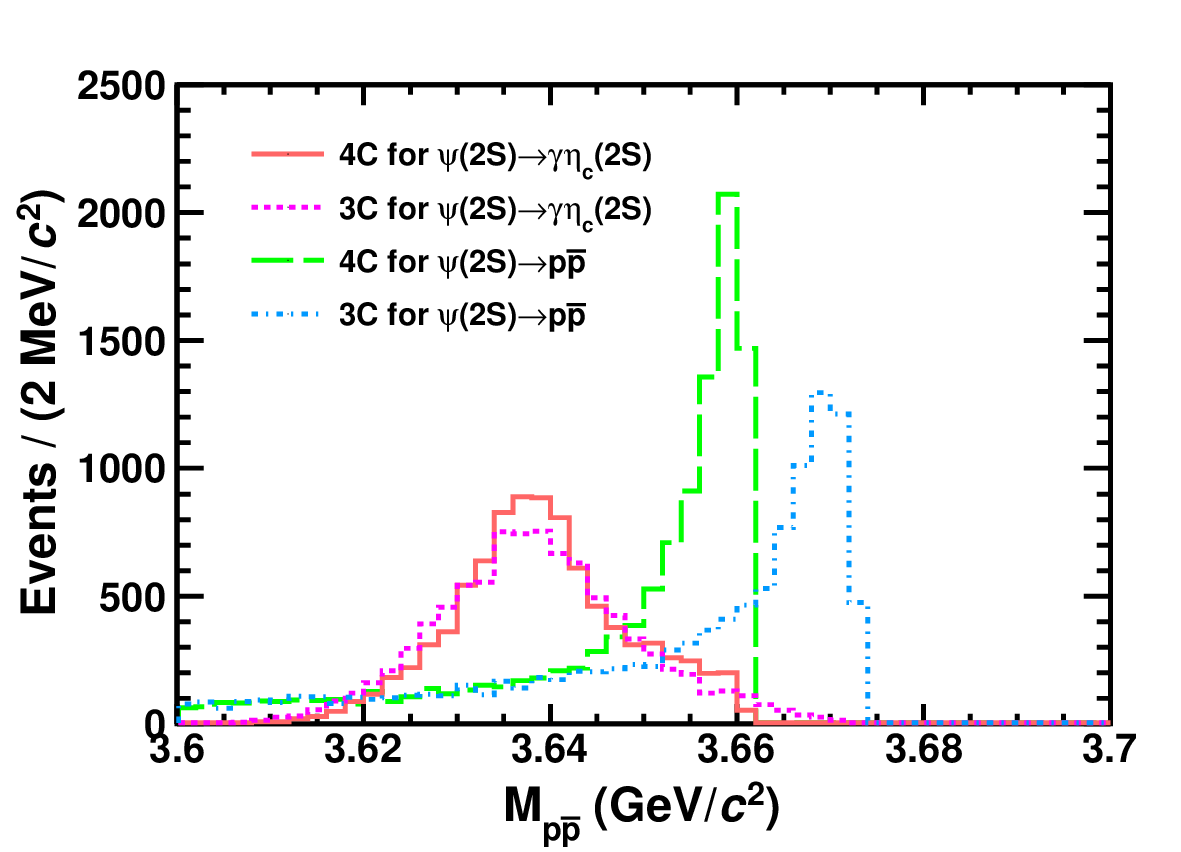}
\end{overpic} \end{center} \caption{The $M_{p\bar{p}}$ distributions
for $\psi(2S) \to \gamma\eta_c(2S), \eta_{c}(2S) \to p\bar{p}$ MC
events with 4C~(red solid) and 3C~(purple dashed) kinematic fits, and
$\psi(2S) \to p\bar{p}$ MC events with 4C~(green dashed) and 3C~(blue
dash-dotted) kinematic fits. The sharp cutoffs of the green and blue
histograms are due to the photon energy threshold of 25~MeV. } 
\label{fig:4c3c} \end{figure}

Events of $\psi(2S) \to p\bar{p}$ accompanied by a fake photon or a FSR
photon can easily pass through the event selection. For the events
with a fake photon, the four-momentum of the proton and anti-proton is
expected to equal to that of $\psi(2S)$. Based on this, a
three-constraint (3C) kinematic fit is performed where the momentum
magnitude of the photon is allowed to float. Figure~\ref{fig:4c3c}
shows the $M_{p\bar{p}}$ distributions from $\psi(2S) \to
\gamma\eta_c(2S), \eta_{c}(2S) \to p\bar{p}$, and $\psi(2S) \to
p\bar{p}$ MC samples after 4C and 3C kinematic fits. The peak of
$\psi(2S) \to p\bar{p}$ is significantly separated from the
$\eta_c(2S)$ signal after the 3C kinematic fit. Therefore, the
$M_{p\bar p}^{3\rm C}$ distribution is used to determine the signal
yield. In addition, $\chi_{4\rm C}^{2}(\gamma p\bar{p})<\chi_{4\rm
C}^{2}(p\bar{p})$ is required to further suppress the background.

The consistency of the FSR photon between MC simulation and data has
been checked using the control sample $J/\psi \to p \bar{p}\gamma_{\rm
FSR}$.  With the same proton and anti-proton selection criteria as for
our signal, the selected numbers of events with a FSR photon are 
$3307\pm 58$ in data and $3200 \pm 57$ in MC, which are consistent with each
other within the statistical uncertainty. Thus, we use the MC
simulation to describe the FSR contribution in our fit process
directly.


\subsection{\boldmath Background of $\psi(2S) \to \pi^{0}p\bar{p}$ }

\begin{figure}[htbp] \begin{center}
\begin{overpic}[width=0.48\textwidth,angle=0]{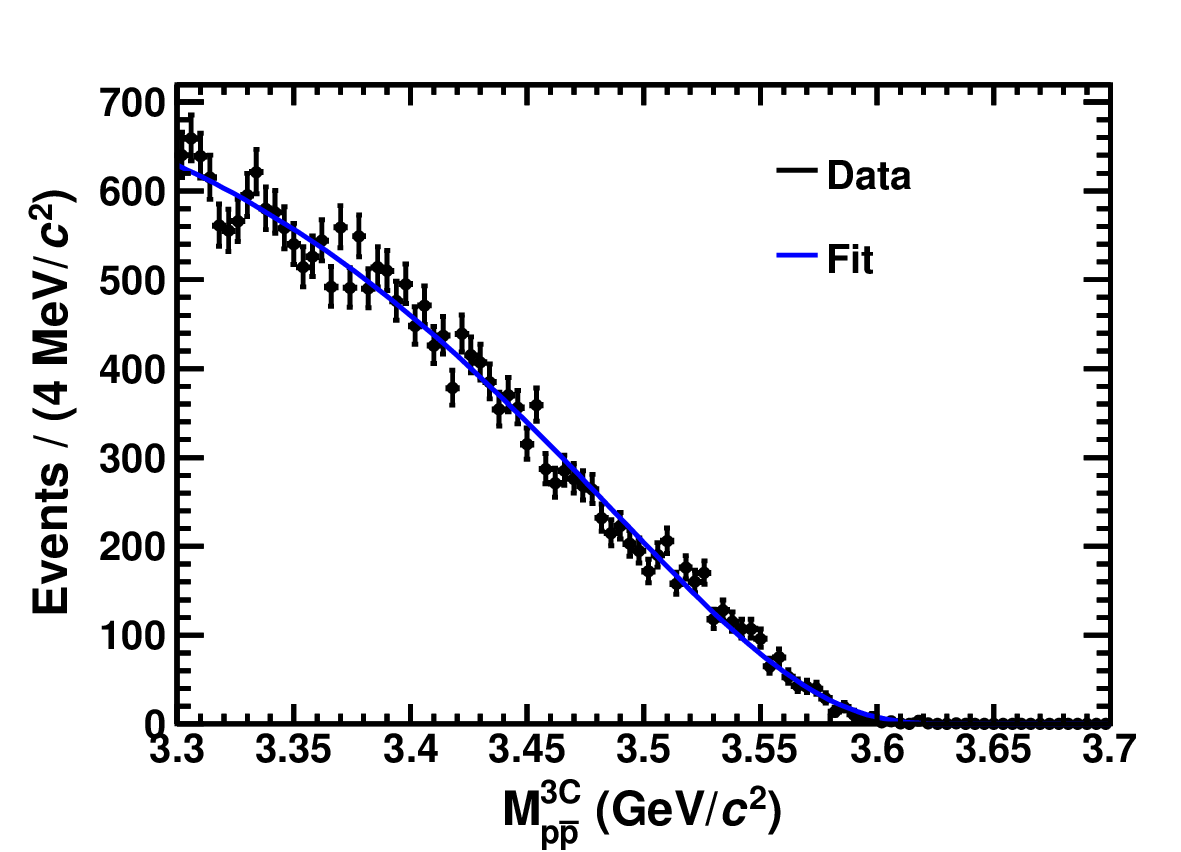}
\end{overpic} \end{center}
\caption{The $M^{3\rm C}_{p \bar{p}}$ distribution.  The black dots
  with error bars are the MC simulated $\psi(2S) \to \pi^{0}p\bar{p}$
  events. The blue solid curve is the fit result with a Novosibirsk
  function.}
\label{fig:pi0pp} \end{figure}

The process of $\psi(2S) \to \pi^{0}p\bar{p}$ can contaminate our
signal if a soft photon is not detected. To estimate this
contribution, we generate corresponding MC samples based on
partial wave analysis results~\cite{pi0pp}. After the event selection,
the distribution of $M^{3\rm C}_{p \bar{p}}$, which is described by a
Novosibirsk function~\cite{Novosibirsk}, is shown in
Fig.~\ref{fig:pi0pp}. There are two solutions for the branching
fraction of $\psi(2S) \to \pi^{0} p\bar{p}$ due to the interference
between the resonance and continuum production, which are
$(133.9\pm11.2\pm2.3)\times10^{-6}$ for constructive interference and
$(183.7\pm13.7\pm3.2)\times10^{-6}$ for destructive interference. We
choose the second one as the nominal value to estimate the number of
background events because it is more consistent with results from
a data-driven method.

\subsection{\boldmath Continuum background }

The continuum production contribution is estimated with a data sample
taken at $\sqrt s=3.65$ GeV. Considering the energy difference between
3.65 and 3.686 GeV, the $M_{p\bar{p}}$ distribution is shifted according to the
transformation: $m \to a(m-m_0)+m_0$, where $m_0=1.877$ GeV/$c^2$ is
the mass threshold of $p\bar{p}$, and the coefficient
$a=(3.686-m_0)/(3.65-m_0)=1.02$. The number of events is scaled based
on the cross sections and luminosities at the two energy points.
the resulting scale factor is calculated to be $f_{\rm continuum} =
\frac{\mathcal{L}_{3.686}}{\mathcal{L}_{3.65}} \cdot
(\frac{3.65}{3.686})^2=9.73$.


\section{\boldmath Branching Fraction Measurement}
\label{sec:bf}

The signal yields are obtained by performing an un-binned maximum
likelihood fit to the $M^{3\rm C}_{p \bar{p}}$ distribution in the
range of $[3.3,3.7]~\rm{GeV}/c^2$, which covers the $\eta_c(2S)$ and
$\chi_{cJ}$ signal regions. The line-shapes of $\eta_c(2S)$ and
$\chi_{cJ}$ are described as \begin{linenomath*} \begin{equation*}
\begin{aligned} (E^3_\gamma \times BW(m;m_0,\Gamma) \times
f_d(E_\gamma) \times \epsilon(m))\otimes \rm DG. \end{aligned}
\end{equation*} \end{linenomath*} Here, $m$ is $M_{p \bar{p}}^{3\rm
C}$, and the first term $E_\gamma^3$ is the PHSP factor, where
$E_\gamma$ is the energy of the transition photon in the rest frame of
$\psi(2S)$, calculated as $E_\gamma =
(m^2_{\psi(2S)}-m^2)/(2m_{\psi(2S)})$, with $m_{\psi(2S)}$ being the
mass of $\psi(2S)$~\cite{PDG}. $BW(m;m_0,\Gamma)$ is the Breit-Wigner
function, with $m_0$ and $\Gamma$ as the masses and widths of
$\eta_c(2S)$ and $\chi_{cJ}$~\cite{PDG}. $f_d (E_\gamma)$ is a damping
factor used to suppress the divergence in the lower side of the mass
spectrum. The form of the damping function used
in the nominal fit, proposed by the KEDR collaboration~\cite{KEDR}, is
taken as $\frac{E_0^2}{E_\gamma E_0+(E_\gamma -E_0)^2}$, where $E_0 =
\frac{m_{\psi(2S)}^2-m_{\eta_c(2S)}^2}{2m_{\psi(2S)}}$ is the most
probable energy of the transition photon. The efficiency curve
$\epsilon(m)$ is based on the PHSP MC sample. We divide
the $M^{3\rm C}_{p \bar{p}}$ distribution into 40 bins, calculate
the efficiency for each bin, and fit these efficiencies to
obtain the curve.  The detector resolution is modeled by a
double-Gaussian~(DG) function. For $\chi_{cJ}$ signals, the parameters
of the DG function are free, while for the $\eta_c(2S)$ signal, the parameters
are extrapolated from the $\chi_{c1}$ and $\chi_{c2}$ parameters
with a first-order polynomial function and are fixed.

In the fitting process, four background components are considered. The
line-shape of $\psi(2S) \to p\bar p$ is
modeled by a Crystal-Ball~(CB)~\cite{cb} function convolved with a DG
function. The parameters of the CB function are fixed based on MC
simulation, while the parameters of the DG function are floated. The
background from $\psi(2S) \to \pi^0 p\bar p$ is described by a
Novosibirsk function, with the yield fixed at $3043\pm55$. The shape
of the non-resonant $\psi(2S) \to \gamma p\bar{p}$ process is
determined by MC simulation, and its magnitude is fixed at $2001\pm45$
according to the PDG branching fraction~\cite{PDG}. Additionally, the
line-shape of the continuum production is fixed, and the number of events is
$243\pm16$.

\begin{figure*}[htbp]
\begin{overpic}[width=0.48\textwidth,angle=0]{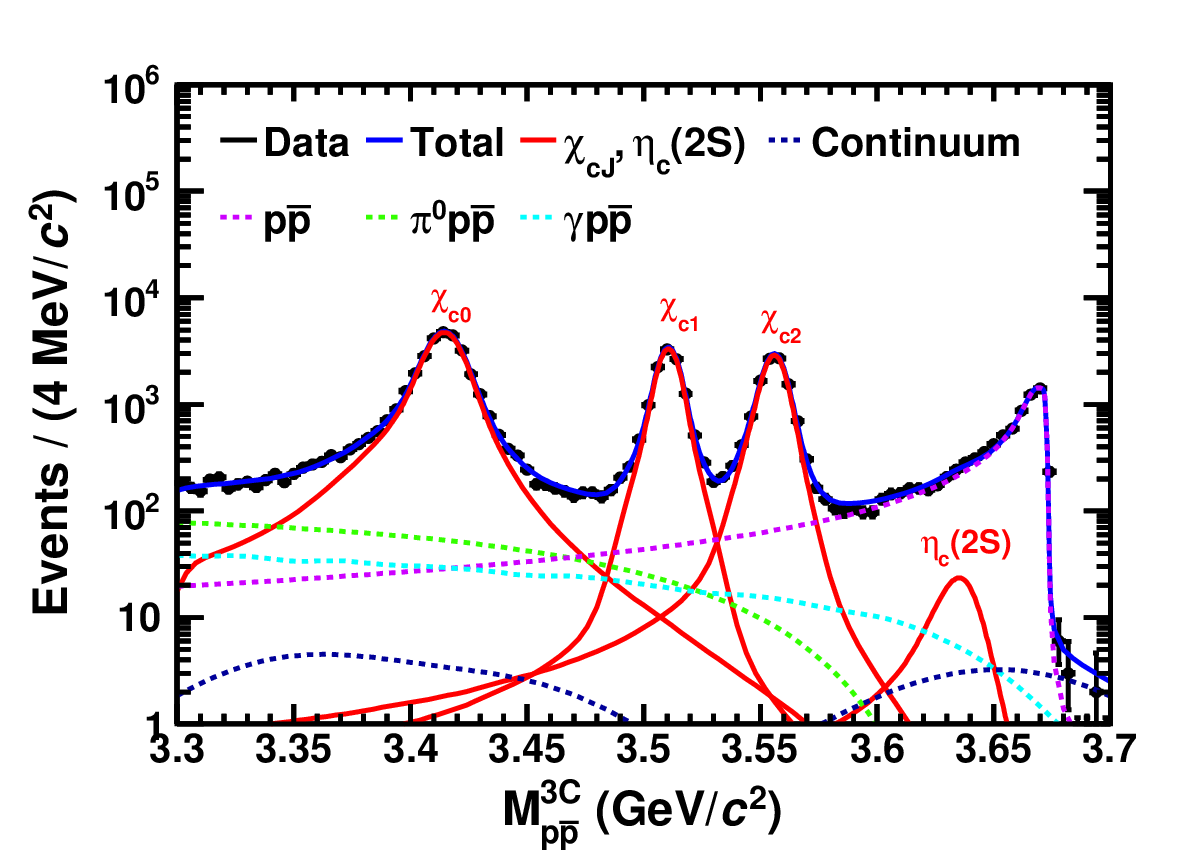}
\end{overpic}
\begin{overpic}[width=0.48\textwidth,angle=0]{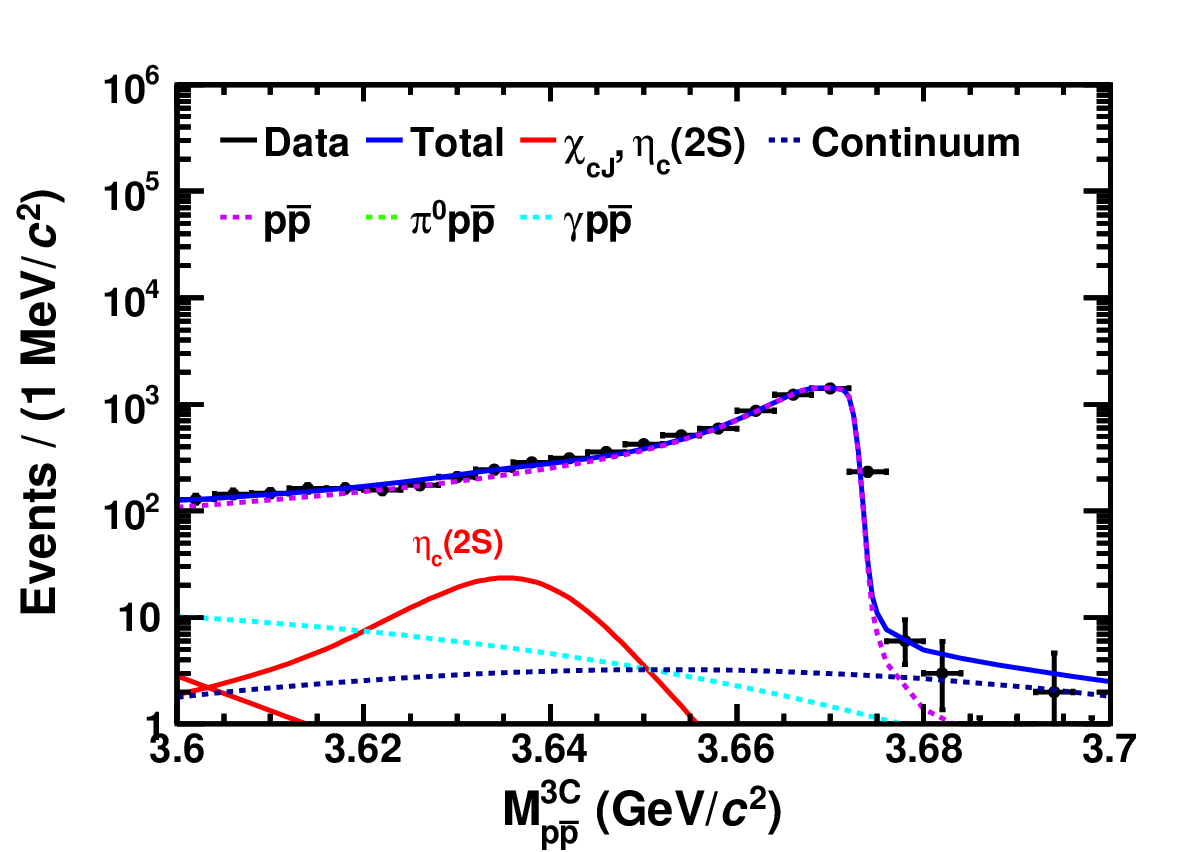}
\end{overpic}

\caption{The $M^{3\rm C}_{p \bar{p}}$ distribution and fit result in
the full fit range (left) and the $\eta_c(2S)$ signal region
(right). The black dots with error bars are data, and the blue-solid
curve is the total fit. The four red-solid lines are the $\chi_{c0}$,
$\chi_{c1}$, $\chi_{c2}$, and $\eta_c(2S)$ signals. The purple, green,
cyan, and dark-blue dashed lines show the background shapes of
$\psi(2S) \to p \bar{p}(\gamma_{\rm FSR})$, $\psi(2S)\to\pi^0 p\bar
p$, $\psi(2S)\to\gamma p\bar{p}$, and the continuum production,
respectively. } \label{fig:fit}

\end{figure*}

Figure~\ref{fig:fit} shows the $M^{3\rm C}_{p \bar{p}}$ distribution
after event selection and the fit results. The left panel shows the
full fit range, while the right panel focuses on the $\eta_c(2S)$ signal
region. The goodness-of-fit is $\chi^2 / \rm{ndf}=39.25/25=1.57$,
where $\rm ndf$ is the number of degrees of freedom.  The branching
fractions of $\chi_{cJ} \to p\bar{p}$ are calculated by
\begin{linenomath*}
\begin{equation*} \begin{aligned}
\mathcal{B}(\chi_{cJ}\to p\bar{p})= \frac{N^{\rm
obs}_{\rm J}}{N_{\psi(2S)}^{\rm tot} \times \mathcal{B}(\psi(2S) \to
\gamma\chi_{cJ}) \times \epsilon^{\rm MC}_{\rm J}}, \end{aligned}
\end{equation*} \end{linenomath*} 
where $N^{\rm obs}_{\rm J}$ are the signal
yields from the fit, $\mathcal{B}(\psi(2S) \to \gamma\chi_{cJ})$ are
taken from PDG~\cite{PDG}, and $\epsilon^{\rm MC}_{\rm J}$ are the
detection efficiencies. The signal yields, detection efficiencies,
and the corresponding numerical results are listed in
Table~\ref{tab:yields}.

The statistical significance of the $\eta_c(2S)$ signal is estimated
to be 2.5\,$\sigma$ by comparing the likelihood values with and
without the signal component. The detection efficiency for $\eta_c(2S)
\to p\bar p$ is $(45.5\pm0.2)\%$, and the signal yield is
$158\pm63$. Dividing by $\mathcal{B}(\psi(2S) \to
\gamma\eta_{c}(2S))$~\cite{br_etacp}, the branching fraction
$\mathcal{B}(\eta_c(2S) \to p\bar{p})$ is determined to be
$(2.46\pm0.98)\times 10^{-4}$, where the uncertainty is
statistical. Using a Bayesian method~\cite{Bayesian}, the upper limit
of the product branching fraction at 90\% C.~L. is determined to be
\begin{linenomath*} \begin{equation*} \begin{aligned}
\mathcal{B}(\psi(2S)\to \gamma\eta_c(2S)) \times
\mathcal{B}(\eta_c(2S) \to p\bar{p}))<2.2\times10^{-7}. \end{aligned}
\end{equation*} \end{linenomath*}

\begin{table*}[!htbp]
	\centering	
	\caption{Signal yields, detection efficiencies, and the
	measured branching fractions of $\chi_{cJ}\to p\bar{p}$, as
	well as the branching fractions from BESIII previous
	measurements~\cite{besiii2013} and the PDG~\cite{PDG}. Here the
	first uncertainties are statistical and the second systematic.
	}
	\label{tab:yields}
	\bgroup
	\def\arraystretch{1.5}
	\setlength{\tabcolsep}{2.5mm} {
    \begin{tabular}{c c c c c c}
	   \hline\hline
    Channel                     &        $N^{\rm obs}_{\rm J}$         &   $\epsilon^{\rm MC}_{\rm J}$(\%)   &               $\mathcal{B}$\,(This work)                         &             BESIII (2013)                                  &            PDG                                \\ \hline 
$\chi_{c0}\to p\bar{p}$   &  $31268 \pm 189$   &        $47.0\pm0.2$                    &  $(2.51 \pm 0.02\pm 0.08)\times10^{-4}$    &  $(2.45\pm 0.08\pm 0.13)\times10^{-4}$    &   $(2.21\pm 0.08)\times10^{-4}$ \\ 
$\chi_{c1}\to p\bar{p}$   &  $11279 \pm 119$   &        $52.2\pm0.2$                    &  $(8.16 \pm 0.09\pm 0.25)\times10^{-5}$    &  $(8.6 \pm 0.5\pm 0.5)\times10^{-5}$         &   $(7.60 \pm 0.34)\times10^{-5}$ \\ 
$\chi_{c2}\to p\bar{p}$   &  $10672 \pm 115$   &        $49.6\pm0.2$                    &  $(8.33 \pm 0.09\pm 0.22)\times10^{-5}$    &  $(8.4 \pm 0.5\pm 0.5)\times10^{-5}$         &   $(7.33 \pm 0.33)\times10^{-5}$ \\
	   \hline\hline
   	\end{tabular} }
   	\egroup
\end{table*}

\section{\boldmath Systematic uncertainties} 

The systematic uncertainties in the branching fraction measurements
are listed in Table~\ref{tab:sys}. The systematic uncertainties are
divided into two parts: the multiplicative and additive terms. The
multiplicative uncertainties include tracking, PID, photon
reconstruction, kinematic fit, generator model, and number of
$\psi(2S)$ events. The additive uncertainties are those related to the
fit process, including the form of the damping function, efficiency curve,
DG parameters, number of $\gamma p\bar{p}$ background events, shape and size
of $\pi^{0}p\bar{p}$ background, and number of continuum
background events. The total systematic uncertainty is obtained by summing
all contributions in quadrature, assuming they are independent.

\begin{table*}[!htbp]
	\centering	
	\caption{Relative systematic uncertainties (in \%) in the
	branching fraction measurements of $\chi_{cJ}\to
	p\bar{p}~(J=0,1,2)$ and the search for $\eta_{c}(2S) \to
	p\bar{p}$. }
	\label{tab:sys}
	\bgroup
	\def\arraystretch{1.5}
	\setlength{\tabcolsep}{3.5mm} {
	\begin{tabular}{l c c c c}
	\hline\hline
Source                                           & $\chi_{c0}\to p\bar{p}$     &       $\chi_{c1}\to p\bar{p}$    &       $\chi_{c2}\to p\bar{p}$    &   $\eta_{c}(2S)\to p\bar{p}$ \\\hline 
Tracking                                        &		$1.10$      &	$0.96$	&	$0.91$	&	$0.8$	\\
PID                                                &		$0.96$	&	$0.99$	&	$1.10$	&	$1.7$	\\
Photon reconstruction                   &		$0.50$	& 	$0.50$	&	$0.50$	&	$1.0$	\\
Kinematic fit                                   &		$0.11$	&	$0.10$	&	$0.10$	&	$0.2$	\\
Generator model                            &		$0.21$	&	$0.19$	&	$0.20$	&      $6.8$ 		\\
Number of $\psi(2S)$ events         &		$0.52$	&	$0.52$	&	$0.52$	&	$0.5$	\\
Quoted branching fractions 		&		$2.04$	&
$2.46$	&	$2.10$	&	$38.5$	\\  \hline
Form of damping function              &		$0.00$	&	$0.01$	&	$0.12$	&	$10.1$	\\
Efficiency curve                       	    &		$0.02$	&	$0.02$	&	$0.04$	&	$1.3$	\\
DG parameters                                     &		$0.00$	&	$0.02$	&	$0.00$	&	$14.6$	\\
Shape of $\psi(2S)\to\pi^{0}p\bar{p}$ background             &		$0.03$	&	$0.05$	&	$0.00$	&	$0.6$	\\
Number of $\psi(2S)\to\pi^{0}p\bar{p}$ background events      &		$1.85$	&	$0.79$	&	$0.15$	&	$6.3$	\\
Number of $\psi(2S)\to\gamma p\bar{p}$ background events     &		$0.79$	&	$0.53$	&	$0.34$	&	$3.8$	\\
Number of continuum background events   &		$0.03$	&	$0.02$
&	$0.01$	&	$0.6$	\\ \hline
Total                                               &		$3.30$	&	$3.07$	&	$2.68$	&	$41.2$ \\
	\hline\hline
   	\end{tabular} }
   	\egroup
\end{table*}

To determine the uncertainties of tracking and PID for the proton,
the uncertainties, given by the efficiency differences between data
and MC control samples as a function of transverse momentum, are
reweighted according to the transverse momentum distributions of the
proton and anti-proton. To cover the momentum range of $p$ and
$\bar{p}$, the control samples $J/\psi \to p\bar{p}$ and $e^{+}e^{-}
\to p\bar{p}$ are used. The uncertainties of tracking are 1.10\%,
0.96\%, 0.91\%, and 0.82\%, while the uncertainties due to PID are
0.96\%, 0.99\%, 1.10\%, and 1.65\% for the $\chi_{c0,1,2}$ and
$\eta_c(2S)$ decays, respectively.

The uncertainty of photon reconstruction for $\chi_{cJ}$ decays in
both the barrel and end-cap regions is determined to be 0.5\%, using a
control sample $e^{+}e^{-} \to \gamma \mu^{+}\mu^{-}$ .  The
energy of transition photon from $\psi(2S) \to \gamma \eta_c(2S)$ is
less than $0.1$ GeV, and the systematic uncertainty is assigned to be 1\%
by using the control samples $J/\psi \to \rho^{0} \pi^{0}$ and
$e^{+}e^{-} \to \gamma \gamma$~\cite{sys_photon2}.

To estimate the uncertainty introduced by the kinematic fit, the helix
parameters are corrected in the MC simulation to reduce the difference
between data and MC events~\cite{besiii2013}. The uncertainties of the
kinematic fit are taken as half of the efficiency differences before and
after the helix parameter correction, which are 0.11\%, 0.10\%, 0.10\%,
and 0.22\% for $\chi_{c0,1,2}$ and $\eta_c(2S)$ decays, respectively.

The helicity angular distributions of the proton in the $\chi_{cJ}$ signal
region are measured from data and described by $1+\alpha
\cos^2\theta_2$~\cite{besiii2013}. The $\chi_{cJ} \to p \bar{p}$
decays are simulated with this $\alpha$ value in the MC sample. By
varying the $\alpha$ value by $\pm1\sigma$, the maximum difference of
MC efficiencies is taken as the uncertainty of the generator
model. For the process of $\eta_c(2S)\to p\bar{p}$, the uncertainty is
estimated by taking $\alpha=1$ and $-1$.

The systematic uncertainty of the efficiency curve $\epsilon(m)$ is
estimated by changing the number of bins to 20, 30, 60, 80, 100,
150, and 200. The maximum difference in the branching fractions from
these efficiency curves is taken as the uncertainty. The uncertainty
caused by the form of damping function is estimated by changing it to
${\rm{exp}}(-E_\gamma ^2/8\beta ^2)$, as used by the CLEO
collaboration~\cite{CLEO}. The parameter $\beta$ is free for
$\chi_{cJ}$ signal and fixed at 65 MeV for the $\eta_c(2S)$ signal. The
uncertainty of the $\eta_c(2S)$ DG parameters is estimated using an
alternative set, where the parameters are the same as the $\chi_{c2}$
signal.

Instead of the Novosibirsk function, an ARGUS function~\cite{argus} is
used to describe the shape of the $\psi(2S) \to \pi^{0}p\bar{p}$
background, and the uncertainty in the number of
$\psi(2S) \to \pi^{0}p\bar{p}$ events is determined by the difference in
signal yield compared to the nominal value.  For the number of
other background events, including non-resonant and continuum
processes, we change them by $\pm1\sigma$, and the maximum difference
of the signal yield is taken as the uncertainty.

The number of $\psi(2S)$ events is determined to be $(27.12\pm0.14)
\times 10^{8}$~\cite{number}, and its uncertainty of 0.52\% is taken
as a systematic uncertainty. The branching fractions of $\psi(2S) \to
\gamma \chi_{cJ}~(J=0,1,2)$ are $(9.79\pm0.20)\%$, $(9.75\pm0.24)\%$,
and $(9.52\pm0.20)\%$~\cite{PDG}, corresponding to the uncertainties
of 2.04\%, 2.46\%, and 2.10\%, respectively. The branching fraction of
$\psi(2S) \to \gamma\eta_{c}(2S)$ is $(5.2\pm0.3(\rm stat)\pm0.5(\rm
syst)^{+1.9}_{-1.4}(\rm extr))\times10^{-4}$~\cite{br_etacp},
corresponding to a 38\% uncertainty.

\begin{figure}[htbp]
\begin{center}
\begin{overpic}[width=0.48\textwidth,angle=0]{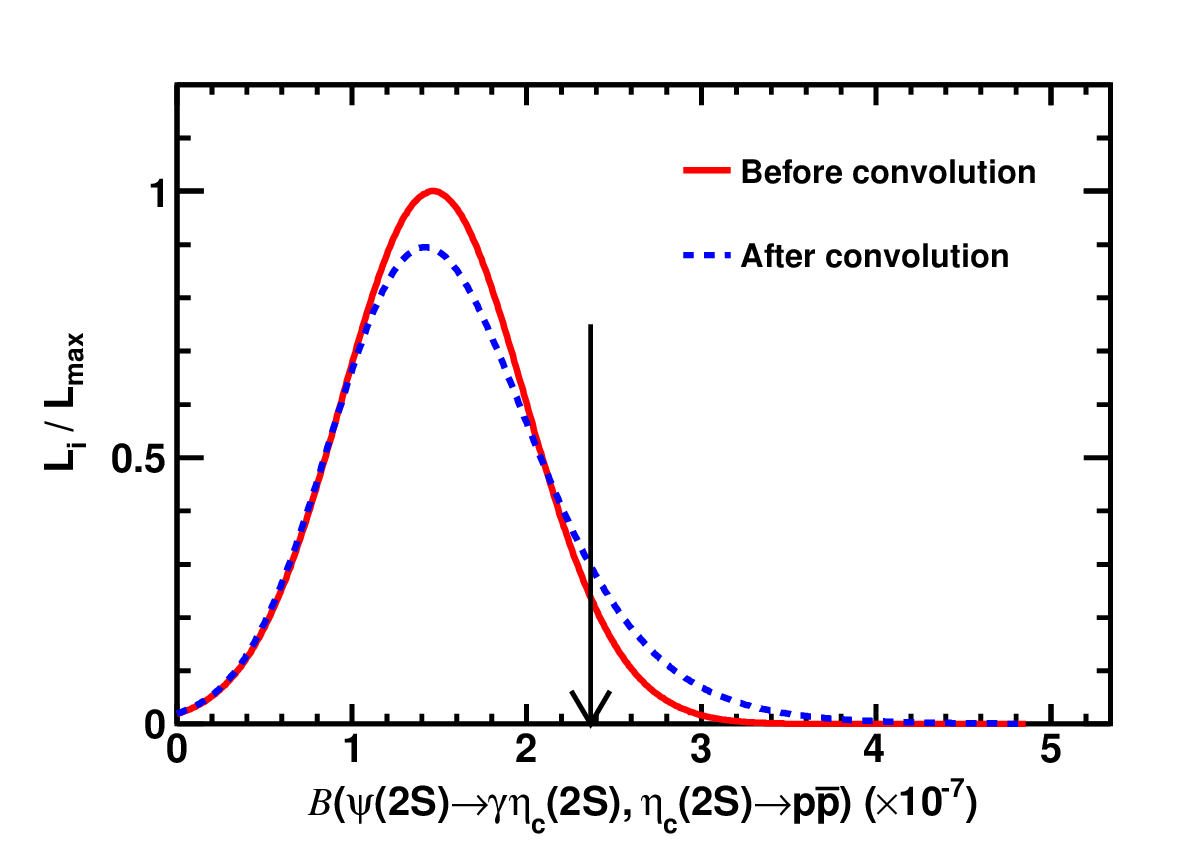}
\end{overpic}
\end{center}
\caption{
The red solid line represents the normalized likelihood distribution incorporating the additive systematic uncertainty, while the blue dashed line further includes the multiplicative systematic uncertainty. The black arrow indicates the upper limit of the product branching fraction at the 90\% C.~L.}

\label{fig:up}
\end{figure}

With fit-related uncertainties considered, the final significance for
$\eta_{c}(2S)$ is conservatively estimated to be $1.7\,\sigma$. It is
the DG function that yields the largest upper limit among the
additive uncertainties. The red solid line in Fig.~\ref{fig:up} shows
the normalized likelihood distribution. This distribution convolved
with a Gaussian function, shown by the blue dashed line, 
shows the effect of multiplicative uncertainty. This process can
be described as \begin{linenomath*} \begin{equation*} \begin{aligned}
L'(x) = \int_{0}^{1} L(x;N_{\rm
sig}\epsilon/\hat{\epsilon})\exp[-\frac{(\epsilon-\hat{\epsilon})}{2\sigma^2_{s}}]d\epsilon,
\end{aligned} \end{equation*} \end{linenomath*} where $L(x;N_{\rm
sig}\epsilon/\hat{\epsilon})$ and $L'(x)$ are the likelihood
distributions before and after incorporating the multiplicative
systematic uncertainty, respectively, $\hat{\epsilon}$ is the nominal
detection efficiency, and $\sigma_{s}$ is the total multiplicative
systematic uncertainty, which is 7.15\% obtained from
Table~\ref{tab:sys}. Taking the systematic uncertainty into account,
the upper limit at the 90\% C.~L. of the product branching fraction is
$2.4\times10^{-7}$.

\section{\boldmath Summary}

Using $(27.12\pm0.14) \times 10^{8}$ $\psi(2S)$ events collected by
the BESIII detector, we search for the decay $\eta_c(2S)\to
p\bar{p}$. A signal with significance of only $1.7\,\sigma$ is
observed. The upper limit of the product branching fraction at the 90\%
C.~L. is determined to be $2.4\times10^{-7}$, which is reduced by an
order of magnitude compared to the previous BESIII measurements.
Dividing by the branching fraction of $\psi(2S)\to
\gamma\eta_c(2S)$~\cite{br_etacp}, the upper limit of the branching
fraction of $\eta_c(2S)\to p\bar{p}$ is calculated to be
$\mathcal{B}(\eta_c(2S)\to p\bar{p})<7.5\times 10^{-4}$.  This result
is consistent with the previous result from the LHCb collaboration,
$\mathcal{B}(\eta_c(2S)\to p\bar{p})=(7.89\pm2.43\pm1.89)\times
10^{-5}$~\cite{LHCb}. 

The branching fractions of $\chi_{cJ}\to p\bar{p}$ are also measured
with improved precision and listed in Table~\ref{tab:yields}, where
the first uncertainties are statistical and the second systematic. Our
results deviate from the PDG~\cite{PDG} values by $2.7\,\sigma$ for
$\chi_{c0}$ and $2.4\,\sigma$ for $\chi_{c2}$, but are consistent with
the previous BESIII measurement~\cite{besiii2013}. The data used in
this analysis incorporates the data from the earlier study. Therefore, this measurement
supersedes that reported in Ref.~\cite{besiii2013}.

\section{\boldmath Acknowledgment}
The BESIII Collaboration thanks the staff of BEPCII and the IHEP computing center for their strong support. This work is supported in part by National Key R\&D Program of China under Contracts Nos. 2020YFA0406300, 2020YFA0406400, 2023YFA1606000; National Natural Science Foundation of China (NSFC) under Contracts Nos. 12275058, 12375070, 11635010, 11735014, 11935015, 11935016, 11935018, 12025502, 12035009, 12035013, 12061131003, 12192260, 12192261, 12192262, 12192263, 12192264, 12192265, 12221005, 12225509, 12235017, 12361141819; the Chinese Academy of Sciences (CAS) Large-Scale Scientific Facility Program; the CAS Center for Excellence in Particle Physics (CCEPP); Joint Large-Scale Scientific Facility Funds of the NSFC and CAS under Contract No. U1832207; 100 Talents Program of CAS; The Institute of Nuclear and Particle Physics (INPAC) and Shanghai Key Laboratory for Particle Physics and Cosmology; German Research Foundation DFG under Contracts Nos. FOR5327, GRK 2149; Istituto Nazionale di Fisica Nucleare, Italy; Knut and Alice Wallenberg Foundation under Contracts Nos. 2021.0174, 2021.0299; Ministry of Development of Turkey under Contract No. DPT2006K-120470; National Research Foundation of Korea under Contract No. NRF-2022R1A2C1092335; National Science and Technology fund of Mongolia; National Science Research and Innovation Fund (NSRF) via the Program Management Unit for Human Resources \& Institutional Development, Research and Innovation of Thailand under Contracts Nos. B16F640076, B50G670107; Polish National Science Centre under Contract No. 2019/35/O/ST2/02907; Swedish Research Council under Contract No. 2019.04595; The Swedish Foundation for International Cooperation in Research and Higher Education under Contract No. CH2018-7756; U. S. Department of Energy under Contract No. DE-FG02-05ER41374.


\end{document}

%% file: authorlist_2024-08-05.tex
\author{
\begin{small}
\begin{center}
M.~Ablikim$^{1}$, M.~N.~Achasov$^{4,c}$, P.~Adlarson$^{76}$, O.~Afedulidis$^{3}$, X.~C.~Ai$^{81}$, R.~Aliberti$^{35}$, A.~Amoroso$^{75A,75C}$, Y.~Bai$^{57}$, O.~Bakina$^{36}$, I.~Balossino$^{29A}$
% ...
\\
\vspace{0.2cm}
(BESIII Collaboration)\\
\vspace{0.2cm} {\it
$^{1}$ Institute of High Energy Physics, Beijing 100049, People's Republic of China\\
% ...
$^{3}$ Bochum  Ruhr-University, D-44780 Bochum, Germany\\
$^{4}$ Budker Institute of Nuclear Physics SB RAS (BINP), Novosibirsk 630090, Russia\\
% ...
$^{29}$ INFN Sezione di Ferrara, (A)INFN Sezione di Ferrara, I-44122, Ferrara, Italy; (B)University of Ferrara,  I-44122, Ferrara, Italy\\
% ...
$^{35}$ Johannes Gutenberg University of Mainz, Johann-Joachim-Becher-Weg 45, D-55099 Mainz, Germany\\
$^{36}$ Joint Institute for Nuclear Research, 141980 Dubna, Moscow region, Russia\\
% ...
$^{57}$ Southeast University, Nanjing 211100, People's Republic of China\\
% ...
$^{75}$ University of Turin and INFN, (A)University of Turin, I-10125, Turin, Italy; (B)University of Eastern Piedmont, I-15121, Alessandria, Italy; (C)INFN, I-10125, Turin, Italy\\
$^{76}$ Uppsala University, Box 516, SE-75120 Uppsala, Sweden\\
% ...
$^{81}$ Zhengzhou University, Zhengzhou 450001, People's Republic of China\\
\vspace{0.2cm}
% ...
$^{c}$ Also at the Novosibirsk State University, Novosibirsk, 630090, Russia\\
% ...
}
\end{center}
\vspace{0.4cm}
\end{small}
}
